\documentclass[runningheads]{llncs}

\usepackage[T1]{fontenc}
\def\doi#1{\href{https://doi.org/\detokenize{#1}}{\url{https://doi.org/\detokenize{#1}}}}
\usepackage{graphicx}
\usepackage{listings}
\usepackage{tkz-graph}
\GraphInit[vstyle = Shade]
\tikzset{
  LabelStyle/.style = { rectangle, rounded corners, draw,
                        minimum width = 2em, fill = yellow!50,
                        text = red, font = \bfseries },
  VertexStyle/.append style = { inner sep=5pt,
                                font = \Large\bfseries},
  EdgeStyle/.append style = {->, bend left} }

\usepackage{lingmacros}

\usepackage{subfigure}
\usepackage{amsfonts}
\usepackage{amssymb}
\usepackage{times}
\usepackage{helvet}
\usepackage{courier}
\usepackage{graphicx,verbatim}
\usepackage{amssymb}
\usepackage{latexsym}
\usepackage{tabularx}
\usepackage{proof}
\usepackage{bussproofs}
\usepackage{amsmath}
\usepackage[toc,page,title,titletoc,header]{appendix}

\usetikzlibrary{decorations.pathmorphing}
\usetikzlibrary{spy,arrows}

\usepackage{braket}

\usepackage{rotating}

\usepackage{wasysym}
\usepackage{tikz}
\usetikzlibrary{shapes.misc,automata,positioning,arrows}
\usepgflibrary{shapes.arrows}

\DeclareMathOperator{\Tr}{Tr}

\newtheorem{Observation}{Observation}
\newtheorem{Definition}{Definition}
\newtheorem{Lemma}{Lemma}
\newtheorem{Theorem}{Theorem}
\newtheorem{Example}{Example}
\newtheorem{Remark}{Remark}
\begin{document}
\title{Distributed Quantum Vote Based on Quantum Logical Operators, a New Battlefield of the Second Quantum Revolution}

% Author Orchid ID: enter ID or remove command
%\newcommand{\orcidauthorA}{0000-0000-000-000X} % Add \orcidA{} behind the author's name
%\newcommand{\orcidauthorB}{0000-0000-000-000X} % Add \orcidB{} behind the author's name

% Authors, for the paper (add full first names)

\author{Xin Sun\inst{1,2}
\and
Feifei He\inst{3}
\and
 Daowen Qiu\inst{4}
 \and
 Piotr Kulicki\inst{1,2}
 \and
 Mirek Sopek\inst{1}
 \and
 Meiyun Guo\inst{5}
 }
\authorrunning{F. Author et al.}

\institute{Quantum Blockchains Inc., 20-027 Lublin, Poland; \email{xin.sun.logic@gmail.com}, \email{\{kulicki,sopek\}@quantumblockchains.io} \and
Department of Foundation of Computer Science, Catholic University of Lublin, 20-950 Lublin, Poland
\email{kulicki@kul.pl}\\
\and
Institute of Logic and Cognition, Sun Yat-sen University, Guangzhou 510275, China; \email{heff5@mail2.sysu.edu.cn}
\and
School of Computer Science, Sun Yat-sen University, Guangzhou 510275, China; \email{issqdw@mail.sysu.edu.cn}
\and
Institute of Logic and Intelligence, Southwest University, Choingqing 400715, China; \email{guomy007@swu.edu.cn}
}
\maketitle

%\Author{Xin Sun$^{1}$, Feifei He$^{2,\ast}$, Daowen Qiu$^3$,  Piotr Kulicki $^{4}$, Mirek Sopek$^{5}$ and Meiyun Guo$^{6}$}

% Affiliations / Addresses (Add [1] after \address if there is only one affiliation.)
%\address{%
%$^{1}$Department of Foundation of Computer Science, Catholic University of Lublin, 20-950 Lublin, Poland;\\
%Quantum Blockchains Inc., 20-027 Lublin, Poland; xin.sun.logic@gmail.com\\
%$^2$Institute of Logic and Cognition, Sun Yat-sen University, Guangzhou 510275, China; heff5@mail2.sysu.edu.cn\\
%$^3$School of Computer Science, Sun Yat-sen University, Guangzhou 510275, China; issqdw@mail.sysu.edu.cn\\
%$^{4}$Department of Foundation of Computer Science, Catholic University of Lublin, 20-950 Lublin, Poland; \\
%Quantum Blockchains Inc., 20-027 Lublin, Poland; piotr.kulicki@kul.pl\\
%$^{5}$Quantum Blockchains Inc., 20-027 Lublin, Poland; sopek@quantumblockchains.io\\
%$^{6}$Institute of Logic and Intelligence, Southwest University, Choingqing 400715, China; guomy007@swu.edu.cn\\
%}

% Contact information of the corresponding author
%\corres{Correspondence: heff5@mail2.sysu.edu.cn %MDPI: it is different from redmine.
%}

% Current address and/or shared authorship

% The commands \thirdnote{} till \eighthnote{} are available for further notes

%\simplesumm{} % Simple summary

%\conference{} % An extended version of a conference paper

% Abstract (Do not insert blank lines, i.e. \\) 

\begin{abstract}
We designed two rules of binary quantum computed vote: Quantum Logical Veto (QLV) and Quantum Logical  Nomination (QLN). The conjunction and disjunction from quantum computational logic are used to define QLV and QLN, respectively. Compared to classical vote, quantum computed vote is fairer, more democratic and has stronger expressive power. Since the advantage of quantum computed vote is neither the speed of computing nor the security of communication, we believe it opens a new battlefield in the second quantum revolution. Compared to other rules of quantum computed vote, QLV and QLN have better scalability. Both QLV and QLN can be implemented by the current technology and the difficulty of implementation does not grow with the increase of the number of voters.

% Keywords
\end{abstract}

% The fields PACS, MSC, and JEL may be left empty or commented out if not applicable
%\PACS{J0101}
%\MSC{}
%\JEL{}

%%%%%%%%%%%%%%%%%%%%%%%%%%%%%%%%%%%%%%%%%%
% Only for the journal Diversity
%\LSID{\url{http://}}

%%%%%%%%%%%%%%%%%%%%%%%%%%%%%%%%%%%%%%%%%%
% Only for the journal Applied Sciences:
%\featuredapplication{Authors are encouraged to provide a concise description of the specific application or a potential application of the work. This section is not mandatory.}
%%%%%%%%%%%%%%%%%%%%%%%%%%%%%%%%%%%%%%%%%%

%%%%%%%%%%%%%%%%%%%%%%%%%%%%%%%%%%%%%%%%%%
% Only for the journal Data:
%\dataset{DOI number or link to the deposited data set in cases where the data set is published or set to be published separately. If the data set is submitted and will be published as a supplement to this paper in the journal Data, this field will be filled by the editors of the journal. In this case, please make sure to submit the data set as a supplement when entering your manuscript into our manuscript editorial system.}

%\datasetlicense{license under which the data set is made available (CC0, CC-BY, CC-BY-SA, CC-BY-NC, etc.)}

%%%%%%%%%%%%%%%%%%%%%%%%%%%%%%%%%%%%%%%%%%
% Only for the journal Toxins
%\keycontribution{The breakthroughs or highlights of the manuscript. Authors can write one or two sentences to describe the most important part of the paper.}

%\setcounter{secnumdepth}{4}
%%%%%%%%%%%%%%%%%%%%%%%%%%%%%%%%%%%%%%%%%%
%\begin{document}
%%%%%%%%%%%%%%%%%%%%%%%%%%%%%%%%%%%%%%%%%%

%%%%%%%%%%%%%%%%%%%%%%%%%%%%%%%%%%%%%%%%%%

\section{Introduction}
% Background
Electronic vote, or e-vote, is a voting process in which ballot casting and counting is computer-aided. Since late 1990s and early 2000s, e-vote has received increasing interest and is widely applied to various situations of decision-making. 
Many voting protocols based on classical cryptography have been developed and successfully
applied in the last two decades \cite{Neff01,Chaum04}. However, the security of protocols based on classical cryptography is based on the unproven complexity of some computational problems, such as the factoring of large numbers. The research in quantum computation shows that quantum computers are able to factor large numbers in a short time, which means that classical protocols based on such algorithms are insecure. To react to the risk posed by forthcoming quantum computers, a number of quantum voting protocols have been developed in the last decade \cite{Hillery06,Vaccaro07,Li08,Horoshko11,Li12,Jiang12,Tian16,Wang16,Rad17,Thapliyal7,Sun19vote,LiJZL21}.
In these works, the ballots are still classical but they are secured by quantum methods. We call this type of vote the quantum secured vote.

While all quantum secured voting protocols have focused on the security problems of voting from the  cryptographic perspective, Bao and Halpern \cite{Bao17} and Sun et al. \cite{SunHSG21} studied quantum vote from a social choice theoretic perspective. They designed voting rules in which ballots are in quantum states and the result of voting is calculated by using quantum operators. We call this type of vote quantum computed vote. An interesting advantage of quantum computed vote is that the quantum analogue of Arrow's Impossibility Theorem \cite{Arrow51} is violated in quantum computed vote.  Arrow's Impossibility Theorem is one of the most influential results in social choice theory. According to the theorem, every voting rule satisfying unanimity and independence of irrelevant alternatives must also satisfy dictatorship, which implies that a fair and democratic voting rule cannot exist. 
The work of Bao and Halpern \cite{Bao17} and Sun et al. \cite{SunHSG21} disproved Arrow's theorem in the quantum setting. Therefore, it provides a theoretic demonstration of the advantage of quantum computed vote: quantum vote is better than classical vote in the sense that it enable the existence of fair and democratic voting rules.

In addition to fairness and democracy, another advantage of quantum computed vote is that it has better expressive power than classical vote. Classical ballots can only be in a definite state like $0$ or $1$, while quantum ballots can be in a superposition of $|0\rangle$ and $|1\rangle$ or mixed state of $|0\rangle\langle 0 | $ and $|1 \rangle\langle 1 | $. In real life, the voters usually have a mixed preference on the proposal to be voted, such as "60\% agree and 40\% disagree". 
 When voters are only allowed to use 0 or 1 to express their preference, it can happen that the result of voting does not truly reveal the aggregation of the preferences of voters. For example, suppose the preference of Alice, Bob and Charlie on a proposal is 0.6 (which means that "60\% agree and 40\% disagree", or Alice will vote for ``agree'' with probability 0.6), 0.6 and 0 respectively. Then they will cast their ballots into classical state ``agree", ``agree" and ``disagree" and the result of voting is ``agree" according to classical majority vote. But intuitively, the overall probability of agree, which is the probability that the majority of the voters vote for ``agree", should be $0.6 \times 0.6= 0.36$. This is because Charlie will never vote for ``agree". In order for the majority of the voters to vote for ``agree", both Alice and Bob must vote for ``agree", which happens with probability $0.6 \times 0.6= 0.36$. Therefore, the result of classical vote does not truly reveal the aggregation of the preferences of voters. This inconsistency is caused by the limited expressive power of classical ballot. On the other hand, the mixed preference like "60\% agree and 40\% disagree" can be described by the quantum state $\sqrt{0.6}|1\rangle+\sqrt{0.4}|0\rangle$ or $ 0.4 |0\rangle\langle 0 |  + 0.6 |1 \rangle\langle 1 | $. Moreover, quantum ballot can also express entangled preference of voters. For example, Alice and Bob together can cast their ballots into state $\frac{|00\rangle + |11 \rangle }{\sqrt{2}}$, which has no analogue in classical voting. Since the main advantages of quantum computed vote is neither the speed of computing nor the security of communication. We believe quantum computed vote opens a new battlefield in the second quantum revolution.

From a practical perspective, the quantum voting rules proposed in Bao and Halpern \cite{Bao17} and Sun et al. \cite{SunHSG21} to disprove Arrow's Impossibility Theorem are too complicated to be realized with the current technology. The number of qubits that are needed to be manipulated in the voting rules is exponential in the growth of the number of voters. Therefore, simpler voting rules are needed. In this paper, we propose two quantum voting rules that have better scalability: quantum logical veto (QLV) and quantum logical nomination (QLN). The number of qubits needed in the two rules is of constant number 3.  The only quantum operation used in QLV/QLN is quantum logical conjunction/disjunction. Both of these operators are relatively simple and have been studied in-depth in the literature of quantum computational logic \cite{Gudder03,Cattaneo04,CattaneoCGL04,LeddaKPG06}. 
Moreover, various voting rules can be constructed by the combination of QLV and QLN without loss of scalability.

The structure of this paper is the following. We introduce elements of background knowledge in Section \ref{Preliminaries}. Then in Section \ref{Quantum voting rules} we introduce our voting rules in detail. We conclude this paper with future work plan in Section \ref{Conclusion and future work}.

\section{Preliminaries}\label{Preliminaries}

Given a Hilbert space $\mathcal{H}=\mathbb{C}^2$, we denote the set of  all density operators on $\mathcal{H}$ by $D(\mathcal{H})$. Our quantum voting rules use two quantum logical operators, quantum AND and quantum OR, for ballot aggregation. The construction of the quantum AND is based on the quantum Toffoli gate \cite{HolikSFGP17}.

\begin{Definition}[Quantum Toffoli gate]

The quantum Toffoli gate is a unitary operator on $\mathbb{C}^{2^3}$:
$$T\ket{x_1, x_2, x_3}=\ket{x_1, x_2, x_1 x_2 \oplus x_3}$$

where $x_i \in \{0,1\}$.

\end{Definition}

\begin{Definition}[Quantum AND operator]

For $\rho,\sigma \in D(\mathbb{C}^2)$, 
$$AND(\rho \otimes \sigma) = \Tr^{1,2}(T(\rho \otimes \sigma \otimes |0\rangle \langle 0| )T^{\dagger}),$$
Here $\Tr^{1,2}$ is the partial trace on the first and the second qubit.

\end{Definition}

\noindent The quantum AND operator is naturally generalized to multiple qubits: $AND(\rho_1  \otimes   \rho_2 \ldots \otimes  \rho_n):= AND(\ldots AND (AND(\rho_1  \otimes  \rho_2 )   \otimes \rho_3) \otimes  \ldots  \otimes \rho_n) $.

Just like in quantum computational logic \cite{Gudder03,Cattaneo04,CattaneoCGL04,LeddaKPG06}, we define the quantum NOT operator by using the Pauli X gate.

\begin{Definition}[Quantum NOT operator]

For $\rho \in D(\mathbb{C}^2)$, $NOT(\rho)=X \rho X^{\dagger}$, where $X$ is the Pauli X operator on a single qubit: $X=\begin{bmatrix} 0 & 1\\1 & 0 \end{bmatrix}$.

\end{Definition}

Now we define the quantum OR operator based on quantum AND and quantum NOT.

\begin{Definition}[Quantum OR operator]

For $\rho,\sigma \in D(\mathbb{C}^2)$, 
$$OR(\rho \otimes \sigma) = NOT(AND(NOT(\rho) \otimes NOT(\sigma)))$$

\end{Definition}

The quantum OR operator is naturally generalized to multiple qubits: $OR(\rho_1   \otimes  \rho_2 \ldots \otimes  \rho_n):= OR(\ldots OR(OR(\rho_1  \otimes  \rho_2 )   \otimes \rho_3 ) \otimes \ldots  \otimes \rho_n)$.

\section{Quantum voting rules}\label{Quantum voting rules}

We design two quantum voting rules: QLV and QLN. In both of them,  we assume there is $1$ proposal to be voted, $m$ voters $\{v_1,\ldots, v_m\}$ and 
 $n$ quantum voting machines $\{M_1,\ldots, M_n\}$.  Every voter's ballot is represented by a density operator of a single qubit. The ballot in state $|0\rangle \langle 0|$ represents ``disagree" and in state $|1\rangle \langle 1|$ represents ``agree". Every quantum voting machine is a small-scale quantum information processor. In QLV (resp. QLN), we assume that the quantum voting machine is able to execute the quantum AND (resp. OR) operator. 

\subsection{Quantum veto}

The one-vote veto is a special type of vote, in which the proposal to be voted will be disagreed as long there is one voter who votes for ``disagree". It has been widely used by many political and economic organizations, among which the most famous is the UN Security Council's permanent
member states group.

There are some work on quantum-secured veto \cite{WuSWCHDX21,WangLYSZ21,mishra21}, in which the ballots are still classical, but they are encrypted by some methods of quantum cryptography in order to ensure some security properties.
 To the best of our knowledge, quantum veto in which ballots are in quantum state has never been studied before.

Our QLV is performed in the following steps:
\begin{enumerate}
\item Every voter $v_i$ sends her/his ballot $\rho_i \in D(\mathbb{C}^2)$ to every quantum voting machine.
    \item Every quantum voting machine $M_j$ calculate $AND(\rho_1 \otimes \dots \otimes \rho_m) = \rho^{j}$.
    \item Every quantum voting machine $M_j$ measures $\rho^{j}$ by the projector $P_1 = |1\rangle \langle 1 |$. It records 1 if the result of measurement is ``yes". It records 0 if the result of measurement is ``no". 
    \item Every quantum voting machine sends its record to every other quantum voting machine. 
    \item Every quantum voting machine reads all the records it has received and outputs ``Agree'' if 
  at least half of the records is 1, otherwise it outputs ``Disagree''.
    
\end{enumerate}

In order to show that QLV indeed satisfies some desirable properties of veto-like voting, we first define the winning probability  of a ballot as follows.

\begin{Definition}[winning probability]

For $\rho \in D(\mathbb{C}^2)$, the winning probability of $\rho$ is $\mathsf{WP}(\rho):= \Tr(P_1 \rho  )$.

\end{Definition}

\begin{Lemma}
For all $\rho,\sigma \in D(\mathbb{C}^2)$, $\Tr^1 \otimes \Tr^2 \otimes \Tr^3 P_1^3 (T(\rho \otimes \sigma \otimes |0\rangle \langle 0| ))T^{\dagger})= \Tr(P_1 \rho) \cdot \Tr(P_1 \sigma) $.
\end{Lemma}

\begin{proof}

We first consider cases where $\rho,\sigma$ ranges over the computational basis. Then we have the following cases:

\begin{enumerate}

\item  $\Tr^1 \otimes \Tr^2 \otimes \Tr^3  P_1^3 (T(|0\rangle \langle 0| \otimes |0\rangle \langle 0| \otimes |0\rangle \langle 0| )T^{\dagger})=   \Tr^1 \otimes \Tr^2 \otimes \Tr^3  P_1^3 (|0\rangle \langle 0| \otimes |0\rangle \langle 0| \otimes |0\rangle \langle 0| )=  \Tr(|0\rangle \langle 0|) \cdot \Tr(|0\rangle \langle 0|) \cdot \Tr P_1(|0\rangle \langle 0|)=1\cdot 1\cdot 0=0= \Tr(P_1 |0\rangle \langle 0|) \cdot \Tr(P_1 |0\rangle \langle 0|)$.

\item  $\Tr^1 \otimes \Tr^2 \otimes \Tr^3  P_1^3 (T(|0\rangle \langle 0| \otimes |1\rangle \langle 1| \otimes |0\rangle \langle 0| )T^{\dagger})=   \Tr^1 \otimes \Tr^2 \otimes \Tr^3  P_1^3 (|0\rangle \langle 0| \otimes |1\rangle \langle 1| \otimes |0\rangle \langle 0| )=  \Tr(|0\rangle \langle 0|) \cdot \Tr(|1\rangle \langle 1|) \cdot \Tr P_1(|0\rangle \langle 0|)=1\cdot 1\cdot 0=0= \Tr(P_1 |0\rangle \langle 0|) \cdot \Tr(P_1 |1\rangle \langle 1|)$.

\item  $\Tr^1 \otimes \Tr^2 \otimes \Tr^3  P_1^3 (T(|1\rangle \langle 1| \otimes |0\rangle \langle 0| \otimes |0\rangle \langle 0| )T^{\dagger})=   \Tr^1 \otimes \Tr^2 \otimes \Tr^3  P_1^3 (|1\rangle \langle 1| \otimes |0\rangle \langle 0| \otimes |0\rangle \langle 0| )=  \Tr(|1\rangle \langle 1|) \cdot \Tr(|0\rangle \langle 0|) \cdot \Tr P_1(|0\rangle \langle 0|)=1\cdot 1\cdot 0=0= \Tr(P_1 |1\rangle \langle 1|) \cdot \Tr(P_1 |0\rangle \langle 0|)$.

\item  $\Tr^1 \otimes \Tr^2 \otimes \Tr^3  P_1^3 (T(|1\rangle \langle 1| \otimes |1\rangle \langle 1| \otimes |0\rangle \langle 0| )T^{\dagger})=   \Tr^1 \otimes \Tr^2 \otimes \Tr^3  P_1^3 (|1\rangle \langle 1| \otimes |1\rangle \langle 1| \otimes |1\rangle \langle 1| )=  \Tr(|1\rangle \langle 1|) \cdot \Tr(|1\rangle \langle 1|) \cdot \Tr P_1(|1\rangle \langle 1|)=1\cdot 1\cdot 1=1= \Tr(P_1 |1\rangle \langle 1|) \cdot \Tr(P_1 |1\rangle \langle 1|)$.
\end{enumerate}

We further note that for all $a,b \in \{0,1\}$ with $a\neq b$, it holds that $\Tr^1 \otimes \Tr^2 \otimes \Tr^3  P_1^3 (T(|a\rangle \langle b| \otimes \sigma \otimes |0\rangle \langle 0| ))T^{\dagger}) = 0 $. This plus the fact that $T$ is a linear operator implies that $\Tr^1 \otimes \Tr^2 \otimes \Tr^3 P_1^3 (T(\rho \otimes \sigma \otimes |0\rangle \langle 0| ))T^{\dagger})= \Tr(P_1 \rho) \cdot \Tr(P_1 \sigma) $ for all $\rho,\sigma \in D(\mathbb{C}^2)$.
\end{proof}

\begin{Lemma}
For $\rho,\sigma \in D(\mathbb{C}^2)$, $\mathsf{WP}(AND(\rho \otimes \sigma) ) =  \mathsf{WP}(\rho) \cdot \mathsf{WP}(\sigma)$.
\end{Lemma}

\begin{proof}
$\mathsf{WP}(AND(\rho \otimes \sigma) ) = \Tr ( P_1 AND(\rho \otimes \sigma) ) = \Tr ( P_1   \Tr^{1,2}(T(\rho \otimes \sigma \otimes |0\rangle \langle 0| ))T^{\dagger})  =  \Tr ( P_1   (\Tr^1 \otimes \Tr^2 \otimes I^3 )(T(\rho \otimes \sigma \otimes |0\rangle \langle 0| ))T^{\dagger})   =     (\Tr^1 \otimes \Tr^2 \otimes \Tr^3 P_1^3 )(T(\rho \otimes \sigma \otimes |0\rangle \langle 0| ))T^{\dagger} = \Tr(P_1 \rho) \cdot \Tr(P_1 \sigma) = \mathsf{WP}(\rho) \cdot \mathsf{WP}(\sigma)$.
\end{proof}

The following theorem states that QLV is indeed a veto-like voting rule.
 \begin{Theorem}
 
 For every quantum voting machine, it records 0 with probability 1 iff at least one voter's ballot is in state $|0\rangle \langle 0|$.

 \end{Theorem}

\begin{proof}

A quantum voting machine records 0 with probability 1 iff it records 1 with probability 0 iff
$\mathsf{WP}(AND(\rho_1 \otimes \dots \otimes \rho_m) ) = 0$ iff $\mathsf{WP}(\rho_1   ) \cdot \ldots \cdot  \mathsf{WP}(\rho_m   )= 0$ iff $\mathsf{WP}(\rho_i   )= 0$ for some voter $v_i$.
\end{proof}

\begin{Remark}
After Step 4 in QLV, every quantum voting machine gets the records of all quantum voting machines. Those records is a collection of 0s and 1s. The collection is a probability distribution on $\{0,1\}$ according to the winning probability of $AND(\rho_1 \otimes \dots \otimes \rho_m)$. Every quantum voting machine has the same collection.

\end{Remark}

\subsection{Quantum nomination}

The quantum nomination is dual to quantum veto, in which the proposal to be voted will be agreed as long as there is one voter who votes for ``agree". Intuitively, this type of voting can be understood as that a candidate is nominated as long as there is one voter who nominates her/him. 
Classical nomination has been widely used in many political and economic elections. Nomination-like vote has also been used in some TV programs. For example, in \textit{The Voice of China}, which is a Chinese reality television singing competition, a contestant get elected in the blind audition phase as long as there is at least one coach who votes for her/him.

To the best of our knowledge, quantum nomination in which ballots are in quantum state has never been studied before. Our QLN operates in the following steps:
\begin{enumerate}
\item Every voter $v_i$ sends her/his ballot $\rho_i$ to every quantum voting machine.
    \item Every quantum voting machine $M_j$ calculates $OR(\rho_1 \otimes \dots \otimes \rho_m) = \rho^{j}$.
    \item Every quantum voting machine measures $\rho^{j}$ using the projector $P_1 = |1\rangle \langle 1 |$. It records 1 if the result of measurement is ``yes". It records 0 if the result of measurement is ``no". 
    \item Every quantum voting machine sends its record to every other voting machine. 
    \item Every quantum voting machine reads all the records it has received and outputs ``Agree'' if 
  at least half of the records is 1. Otherwise it outputs ``Disagree''.
\end{enumerate}

The following lemmas and theorem demonstrate that quantum nomination is indeed a voting rule for nomination-like vote. 

\begin{Lemma}
For $\rho \in D(\mathbb{C}^2)$, $\mathsf{WP}(NOT(\rho)  ) =  1- \mathsf{WP}(\rho) $.
\end{Lemma}

\begin{proof}
$\mathsf{WP}(NOT(\rho)  ) = \mathsf{WP}(X(\rho)X^{\dagger}  )= \mathsf{WP}(X(\rho)X  ) = \Tr (P_1 X(\rho)X )  = \Tr (XP_1 X(\rho) )$. By simple calculation we have $P_0:= |0\rangle \langle 0| = XP_1 X $ and $P_0+P_1 =I$. Therefore, $\Tr (XP_1 X(\rho) )= \Tr (P_0 (\rho) ) =  \Tr ( ( I-P_1 ) \rho ) = \Tr (  \rho -  P_1  \rho )= \Tr(\rho) - \Tr(P_1 \rho) = 1- \mathsf{WP}(\rho)$.
\end{proof}

\begin{Lemma}
For $\rho,\sigma \in D(\mathbb{C}^2)$, $\mathsf{WP}(OR(\rho \otimes \sigma) ) =  \mathsf{WP}(\rho)  + \mathsf{WP}(\sigma) - \mathsf{WP}(\rho)  \cdot \mathsf{WP}(\sigma)$.
\end{Lemma}

\begin{proof}

 $\mathsf{WP}(OR(\rho \otimes \sigma) ) =  \mathsf{WP}(NOT(AND(NOT(\rho) \otimes NOT(\sigma))) )
= 1- \mathsf{WP}(AND(NOT(\rho) \otimes NOT(\sigma) )= 1- \mathsf{WP}(NOT(\rho)) \cdot   \mathsf{WP}(NOT(\sigma))= 1- (1- \mathsf{WP}(\rho))\cdot (1- \mathsf{WP}(\sigma)  )= 1- (1-  \mathsf{WP}(\rho) -\mathsf{WP}(\sigma ) + \mathsf{WP}(\rho) \cdot  \mathsf{WP}(\sigma ) )
 =\mathsf{WP}(\rho)  + \mathsf{WP}(\sigma) - \mathsf{WP}(\rho)  \cdot \mathsf{WP}(\sigma)$.
\end{proof}

 \begin{Theorem}
 
 For every quantum voting machine, it records 1 with probability 1 iff at least one voter's ballot is in state $|1\rangle \langle 1|$.

 \end{Theorem}
 
 \begin{proof}
 We only prove cases in which there are only two voters. The case with multiple voters can be generalized straightforwardly.\\
$(\Leftarrow)$ $\mathsf{WP}(OR(|1\rangle \langle 1| \otimes \sigma) )= \mathsf{WP}(|1\rangle \langle 1|)  + \mathsf{WP}(\sigma) - \mathsf{WP}(|1\rangle \langle 1|)  \cdot \mathsf{WP}(\sigma)=1+ \mathsf{WP}(\sigma) -\mathsf{WP}(\sigma)=1$. The case in which $\sigma = |1\rangle \langle 1|$ is similar.\\
 $(\Rightarrow )$ Assume $\sigma $ is not state $ |1\rangle \langle 1|$ and $\rho $ is not state $ |1\rangle \langle 1|$. Then $\mathsf{WP} (\sigma )<1$ and $\mathsf{WP}(\rho)<1$. Then $\mathsf{WP}(OR(\rho \otimes \sigma) ) =  \mathsf{WP}(\rho)  + \mathsf{WP}(\sigma) - \mathsf{WP}(\rho)  \cdot \mathsf{WP}(\sigma) = \mathsf{WP}(\rho) (1- \mathsf{WP}(\sigma)) +  \mathsf{WP}(\sigma) < 1- \mathsf{WP}(\sigma) +  \mathsf{WP}(\sigma) =1$.
 \end{proof}

\subsection{Extension and Application}

In this subsection we use AND and OR to build other voting rules and study some mathematical properties of the quantum computed vote.

\subsubsection{Logical formulas and voting rules}

The essential feature of QLV and QLN is determined by the logical operators they use. It turns out that different quantum voting rules can be defined by different combinations of logical operators. We illustrate some of them in the following examples.

\begin{Example}[role-weighted vote]

Suppose $v_1$ is a professor of quantum information, $v_2$ and $v_3$ are two associate professors of quantum information. Then the following formula determines a voting rule based on the voters' roles:

\begin{center}

$OR(\rho_1 \otimes (AND(\rho_2 \otimes \rho_3)))$.

\end{center}
 
 \noindent
According to the above formula, as long as the professor votes for ``agree'', the proposal will be agreed. Otherwise both of the two associate professors  needs to vote for ``agree'' in order for the proposal to be agreed.

\end{Example}

\begin{Example}[majority vote]

Majority vote is probably the most popular voting rule in our society. The quantum majority vote for three voters can be determined by the following formula:

\begin{center}
$OR((AND(\rho_1 \otimes \rho_2)) \otimes (AND(\rho_2 \otimes \rho_3)) \otimes (AND(\rho_1 \otimes \rho_3))  )$.

\end{center}
 
 \noindent
 Indeed, if at least two voters vote for ``agree'', then the proposal will be agreed according to the above formula. On the other hand, if the preference of the voters is 0.6, 0.6 and 0 respectively, then they can set $\rho_1 =  \rho_2  = 0.4 |0\rangle\langle 0 |  + 0.6 |1 \rangle\langle 1 | $ and $\rho_3= |0\rangle\langle 0 | $. Then $\mathsf{WP}(  OR((AND(\rho_1 \otimes \rho_2)) \otimes (AND(\rho_2 \otimes \rho_3)) \otimes (AND(\rho_1 \otimes \rho_3))  ) = 0.36$.

\end{Example}

\subsubsection{Embedding probabilistic ballot into quantum ballot}

Let $r\in [0,1]$, $| \theta_r \rangle =  \sqrt{1-r} |0\rangle +\sqrt{r} |1\rangle$ and  $\Theta_r:= | \theta_r \rangle \langle  \theta_r | $. Then $\mathsf{WP}(\Theta_r) = r$ and we call the quantum ballot $\Theta_r$ a canonical representation of the probabilistic ballot $r$. Therefore, if a voter's preference is $r$, then 
he can set his ballot into the pure state  $\Theta_r$ to represent his preference. In this way all probabilistic ballot $r$ can be represented by a quantum ballot $\Theta_r$.

Nevertheless, not all quantum ballot can be represented by probabilistic ballot. For example, ballots in the entangled state $\frac{|00\rangle + |11 \rangle }{\sqrt{2}}$ cannot be represented by probabilistic ballots. We demonstrate this fact by the following observations.

\begin{Observation}
If two quantum ballots in state $\frac{|00\rangle + |11 \rangle }{\sqrt{2}}$ are submitted to QLV, then the proposal will be agreed with probability $\frac{1}{2}$. The same probability will appear when they are submitted to QLN.
\end{Observation}

\begin{Observation}
There are no probabilistic ballot $x,y \in [0,1]$ such that when they are submitted to veto/nomination, the proposal will be agreed with probability $\frac{1}{2}$.
\end{Observation}

\begin{proof}
Suppose $x,y \in [0,1]$ are two probabilistic ballots and they produce  probability $\frac{1}{2}$ in both veto and nomination. Then $xy =\frac{1}{2}$ and $x +y - xy = \frac{1}{2}$. Hence $x+y=1$. Then we know $x(1-x)=\frac{1}{2}$, $x^2 -x +\frac{1}{2}=0$. But there is no real $x$ satisfies $x^2 -x +\frac{1}{2}=0$.
\end{proof}

In fact, two quantum ballots $\rho_1, \rho_2$ in states $\rho_1 = \frac{1+i}{2} |0\rangle + \frac{1-i}{2} |1\rangle$ and $\rho_2 = \frac{1-i}{2} |0\rangle + \frac{1+i}{2} |1\rangle$ also produce the same probability in QLV and QLN as the quantum ballots in state $\frac{|00\rangle + |11 \rangle }{\sqrt{2}}$. Therefore, there even exist non-entangled quantum ballots which cannot be represented by probabilistic ballots.

\section{Conclusion and future work}\label{Conclusion and future work}

We have designed two rules of binary quantum computed vote: QLV and QLN. In both of them ballots are cast into quantum states. The conjunction and disjunction from quantum computational logic are used to define quantum veto and quantum nomination, respectively. Compared to other rules of quantum computed vote, QLV and QLN have advantages in scalability. Both of them can be physically realized by the current technology and the difficulty of physical realization does not grow with the increase of the number of voters. They can also be combined to define other interesting and useful quantum voting rules without loss of scalability.

In the future, we will be interested in the physical realization of quantum veto and nomination. For example, an ion trap quantum computer is a good candidate because the realization of the quantum Toffoli gate with trapped ions has been successful since 2009 \cite{Monz09}. We also plan to study quantum veto and nomination in the situation where some quantum voting machines suffer from faulty behaviour such as crash failure or Byzantine failure. In these situations we will use quantum blockchain \cite{Sun19blockchain,Sun19vote} as a platform to execute quantum veto and nomination.

%\section*{Author Contribution}

%Conceptualization, X.S.; Methodology, X.S., P.K. and M.G.; Validation, D.Q. and M.S.; Formal analysis, X.S. and D.Q.; Writing—original draft preparation, H.F.; Writing—review and editing, X.S., D.Q., M.S. and P.K.; All authors have read and agreed to the published version of the manuscript.

\section*{Funding}
The project is funded by the Minister of Education and Science within the program under
the name ``Regional Initiative of Excellence'' in 2019-2022, project
number: 028/RID/2018/19, to the amount: 11,742,500 PLN  and by Polish Agency for Enterprise Development in 2021-2022, project number POPW.01.01.02-06-0031/21.

%\section*{Institutional Review Board Statement}
%Not applicable.

%\section*{Informed Consent Statement}
%Not applicable.

%\section*{Data Availability}
%No datasets were generated or analysed during the current study.

%\section*{Competing Interests}
%The authors declare that there are no competing interests.

%\bibliographystyle{plain}
%\bibliography{generic}

% The following MDPI journals use author-date citation: Arts, Econometrics, Economies, Genealogy, Humanities, IJFS, JRFM, Laws, Religions, Risks, Social Sciences. For those journals, please follow the formatting guidelines on http://www.mdpi.com/authors/references
% To cite two works by the same author: \citeauthor{ref-journal-1a} (\citeyear{ref-journal-1a}, \citeyear{ref-journal-1b}). This produces: Whittaker (1967, 1975)
% To cite two works by the same author with specific pages: \citeauthor{ref-journal-3a} (\citeyear{ref-journal-3a}, p. 328; \citeyear{ref-journal-3b}, p.475). This produces: Wong (1999, p. 328; 2000, p. 475)

%=====================================
% References, variant B: external bibliography
%=====================================
%\externalbibliography{yes}
%\bibliography{your_external_BibTeX_file}

\begin{thebibliography}{10}

\bibitem{Arrow51}
Kenneth~J. Arrow.
\newblock {\em Social Choice and Individual Values}.
\newblock John Wiley and Sons, 1951.

\bibitem{Bao17}
Ning Bao and Nicole Yunger~Halpern.
\newblock Quantum voting and violation of arrow's impossibility theorem.
\newblock {\em Phys. Rev. A}, 95:062306, Jun 2017.

\bibitem{CattaneoCGL04}
Gianpiero Cattaneo, Maria Luisa~Dalla Chiara, Roberto Giuntini, and Roberto
  Leporini.
\newblock Quantum computational structures.
\newblock {\em Mathematica Slovaca}, 54(1):87--108, 2004.

\bibitem{Cattaneo04}
Gianpiero Cattaneo, Maria Luisa~Dalla Chiara, Roberto Giuntini, and Roberto
  Leporini.
\newblock An unsharp logic from quantum computation.
\newblock {\em International Journal of Theoretical Physics}, 43(7):1803--1817,
  Aug 2004.

\bibitem{Chaum04}
D.~Chaum.
\newblock Secret-ballot receipts: True voter-verifiable elections.
\newblock {\em IEEE Security Privacy}, 2(1):38--47, 2004.

\bibitem{Gudder03}
S.~Gudder.
\newblock Quantum computational logic.
\newblock {\em International Journal of Theoretical Physics}, 42(1):39--47, Jan
  2003.

\bibitem{Hillery06}
Mark Hillery, Mario Ziman, Vladimir Buzek, and Martina Bielikova.
\newblock Towards quantum-based privacy and voting.
\newblock {\em Physics Letters A}, 349(1):75 -- 81, 2006.

\bibitem{HolikSFGP17}
Federico Holik, Giuseppe Sergioli, Hector Freytes, Roberto Giuntini, and Angel
  Plastino.
\newblock Toffoli gate and quantum correlations: a geometrical approach.
\newblock {\em Quantum Inf. Process.}, 16(2):55, 2017.

\bibitem{Horoshko11}
Dmitri Horoshko and Sergei Kilin.
\newblock Quantum anonymous voting with anonymity check.
\newblock {\em Physics Letters A}, (375):1172--1175, 2011.

\bibitem{Jiang12}
Lang Jiang, Guangqiang He, Ding Nie, Jin Xiong, and Guihua Zeng.
\newblock Quantum anonymous voting for continuous variables.
\newblock {\em Phys. Rev. A}, 85:042309, Apr 2012.

\bibitem{LeddaKPG06}
Antonio Ledda, Martinvaldo Konig, Francesco Paoli, and Roberto Giuntini.
\newblock Mv-algebras and quantum computation.
\newblock {\em Studia Logica}, 82(2):245--270, 2006.

\bibitem{Li08}
Yuan Li and Guihua Zeng.
\newblock Quantum anonymous voting systems based on entangled state.
\newblock {\em Optical Review}, 15(5):219--223, Sep 2008.

\bibitem{Li12}
Yuan Li and Guihua Zeng.
\newblock Anonymous quantum network voting scheme.
\newblock {\em Optical Review}, 19(3):121--124, May 2012.

\bibitem{LiJZL21}
Yue{-}Ran Li, Dong{-}Huan Jiang, Yong{-}Hua Zhang, and Xiang{-}Qian Liang.
\newblock A quantum voting protocol using single-particle states.
\newblock {\em Quantum Inf. Process.}, 20(3):1--17, 2021.

\bibitem{mishra21}
Sandeep Mishra, Kishore Thapliyal, Abhishek Parakh, and Anirban Pathak.
\newblock Quantum anonymous veto: A set of new protocols, 2021.

\bibitem{Monz09}
T.~Monz, K.~Kim, W.~H\"ansel, M.~Riebe, A.~S. Villar, P.~Schindler, M.~Chwalla,
  M.~Hennrich, and R.~Blatt.
\newblock Realization of the quantum toffoli gate with trapped ions.
\newblock {\em Phys. Rev. Lett.}, 102:040501, Jan 2009.

\bibitem{Neff01}
C.~Andrew Neff.
\newblock A verifiable secret shuffle and its application to e-voting.
\newblock In Michael~K. Reiter and Pierangela Samarati, editors, {\em {CCS}
  2001, Proceedings of the 8th {ACM} Conference on Computer and Communications
  Security, Philadelphia, Pennsylvania, USA, November 6-8, 2001}, pages
  116--125. {ACM}, 2001.

\bibitem{Rad17}
Soroush~Rafiee Rad, Elahe Shirinkalam, and Sonja Smets.
\newblock A logical analysis of quantum voting protocols.
\newblock {\em International Journal of Theoretical Physics},
  56(12):3991--4003, Dec 2017.

\bibitem{SunHSG21}
Xin Sun, Feifei He, Mirek Sopek, and Meiyun Guo.
\newblock Schr{\"{o}}dinger's ballot: Quantum information and the violation of
  arrow's impossibility theorem.
\newblock {\em Entropy}, 23(8):1083, 2021.

\bibitem{Sun19blockchain}
Xin Sun, Mirek Sopek, Quanlong Wang, and Piotr Kulicki.
\newblock Towards quantum-secured permissioned blockchain: Signature,
  consensus, and logic.
\newblock {\em Entropy}, 21(9), 2019.

\bibitem{Sun19vote}
Xin Sun, Quanlong Wang, Piotr Kulicki, and Mirek Sopek.
\newblock A simple voting protocol on quantum blockchain.
\newblock {\em International Journal of Theoretical Physics}, 58(1):275--281,
  Jan 2019.

\bibitem{Thapliyal7}
Kishore Thapliyal, Rishi~Dutt Sharma, and Anirban Pathak.
\newblock Protocols for quantum binary voting.
\newblock {\em International Journal of Quantum Information}, 15(01):1750007,
  2017.

\bibitem{Tian16}
Juan-Hong Tian, Jian-Zhong Zhang, and Yan-Ping Li.
\newblock A voting protocol based on the controlled quantum operation
  teleportation.
\newblock {\em International Journal of Theoretical Physics}, 55(5):2303--2310,
  May 2016.

\bibitem{Vaccaro07}
J.~A. Vaccaro, Joseph Spring, and Anthony Chefles.
\newblock Quantum protocols for anonymous voting and surveying.
\newblock {\em Phys. Rev. A}, 75:012333, Jan 2007.

\bibitem{WangLYSZ21}
Qingle Wang, Yuancheng Li, Chaohang Yu, Runhua Shi, and Zhichao Zhang.
\newblock Quantum-based anonymity and secure veto.
\newblock {\em Quantum Inf. Process.}, 20(3):85, 2021.

\bibitem{Wang16}
Qingle Wang, Chaohua Yu, Fei Gao, Haoyu Qi, and Qiaoyan Wen.
\newblock Self-tallying quantum anonymous voting.
\newblock {\em Phys. Rev. A}, 94:022333, Aug 2016.

\bibitem{WuSWCHDX21}
Songyang Wu, Wenqi Sun, Qingle Wang, Ronghua Che, Meng Hu, Zhiguo Ding, and Xue
  Xue.
\newblock A secure quantum protocol for anonymous one-vote veto voting.
\newblock {\em {IEEE} Access}, 9:146841--146849, 2021.

\end{thebibliography}

%%%%%%%%%%%%%%%%%%%%%%%%%%%%%%%%%%%%%%%%%%
%% optional

%% for journal Sci
%\reviewreports{\\
%Reviewer 1 comments and authors’ response\\
%Reviewer 2 comments and authors’ response\\
%Reviewer 3 comments and authors’ response
%}

%%%%%%%%%%%%%%%%%%%%%%%%%%%%%%%%%%%%%%%%%%
\end{document}